\documentclass[showpacs,amsmath,amssymb,twocolumn,prl,superscriptaddress]{revtex4-1}
\usepackage{amssymb}
\usepackage[dvips]{graphicx}
\usepackage{enumerate}
\usepackage{epsfig}
\usepackage{subfigure}
\usepackage{xcolor}
\usepackage[T1]{fontenc}
\usepackage{fullpage}
\usepackage{amsthm,amsfonts,amssymb,amscd,mathrsfs,xspace,framed}
\usepackage{amsmath}
\usepackage{color}
\usepackage{setspace}
\usepackage{url}
\usepackage{wrapfig}
\usepackage{enumitem}
\begin{document}

%\title{Field Implementation of a Quantum-Enhanced Network with Squeezed-Light-Based Threat Detection}

\title{Quantum‑enhanced physical‑layer threat detection in metropolitan-scale fiber networks}

%\title{Quantum-squeezing-enhanced threat detection in metropolitan-scale fiber networks}

\author{Yung-Cheng Kao}
\affiliation{Chandra Department of Electrical and Computer Engineering, The University of Texas at Austin, Austin, Texas 78758, USA}
\author{Siddharth Pal}
\affiliation{RTX BBN Technologies, Cambridge, Massachusetts 02138, USA}
\author{Alex Forencich}
\affiliation{University of California, San Diego, CA 92093 USA}
\author{Dylan Cirimelli-Low}
\affiliation{RTX BBN Technologies, Cambridge, Massachusetts 02138, USA}
\author{Chaohan Cui}
\affiliation{Department of Electrical and Computer Engineering, University of Maryland, College Park, Maryland 20742, USA}
\author{Jack Postlewaite}
\affiliation{Department of Electrical and Computer Engineering, University of Maryland, College Park, Maryland 20742, USA}
\author{Pao-Kang Chen}
\affiliation{Chandra Department of Electrical and Computer Engineering, The University of Texas at Austin, Austin, Texas 78758, USA}
\author{Nicola Alic}
\affiliation{University of California, San Diego, CA 92093 USA}
\author{Saikat Guha}
\affiliation{Department of Electrical and Computer Engineering, University of Maryland, College Park, Maryland 20742, USA}
\author{Prithwish Basu}
\email{prithwish.basu@rtx.com}
\affiliation{RTX BBN Technologies, Cambridge, Massachusetts 02138, USA}
\author{Linran Fan}
\email{linran.fan@utexas.edu}
\affiliation{Chandra Department of Electrical and Computer Engineering, The University of Texas at Austin, Austin, Texas 78758, USA}

\maketitle

%% Abstract text start
\textbf{
Network security is widely recognized as a key application of quantum technology. However, its large-scale deployment is hindered by the need for tight coordination between fundamentally different quantum and classical processing steps in conventional protocols. This requirement introduces strong cross-layer interdependencies that conflict with the modular, layered architectures enabling scalability in modern communication networks.
Here, we present an alternative strategy that confines all quantum interventions to the physical layer and remains transparently compatible with existing network abstractions.
This is achieved by directly embedding quantum features and classical information within the same optical field using bright squeezed light. Physical-layer signals are analyzed using a cumulative sum (CUSUM) method to enable quantum-enhanced threat detection.
We validate the practicality of this approach through field deployment over a metropolitan-scale fiber network and further demonstrate network-level security functionalities enabled by physical-layer quantum-enhanced thread detection.
These results establish a practical, scalable framework for seamlessly integrating quantum-enhanced security into large-scale communication infrastructure.
}

%% Main text start
\begin{figure*}
\centering
\includegraphics[width=0.98\textwidth]{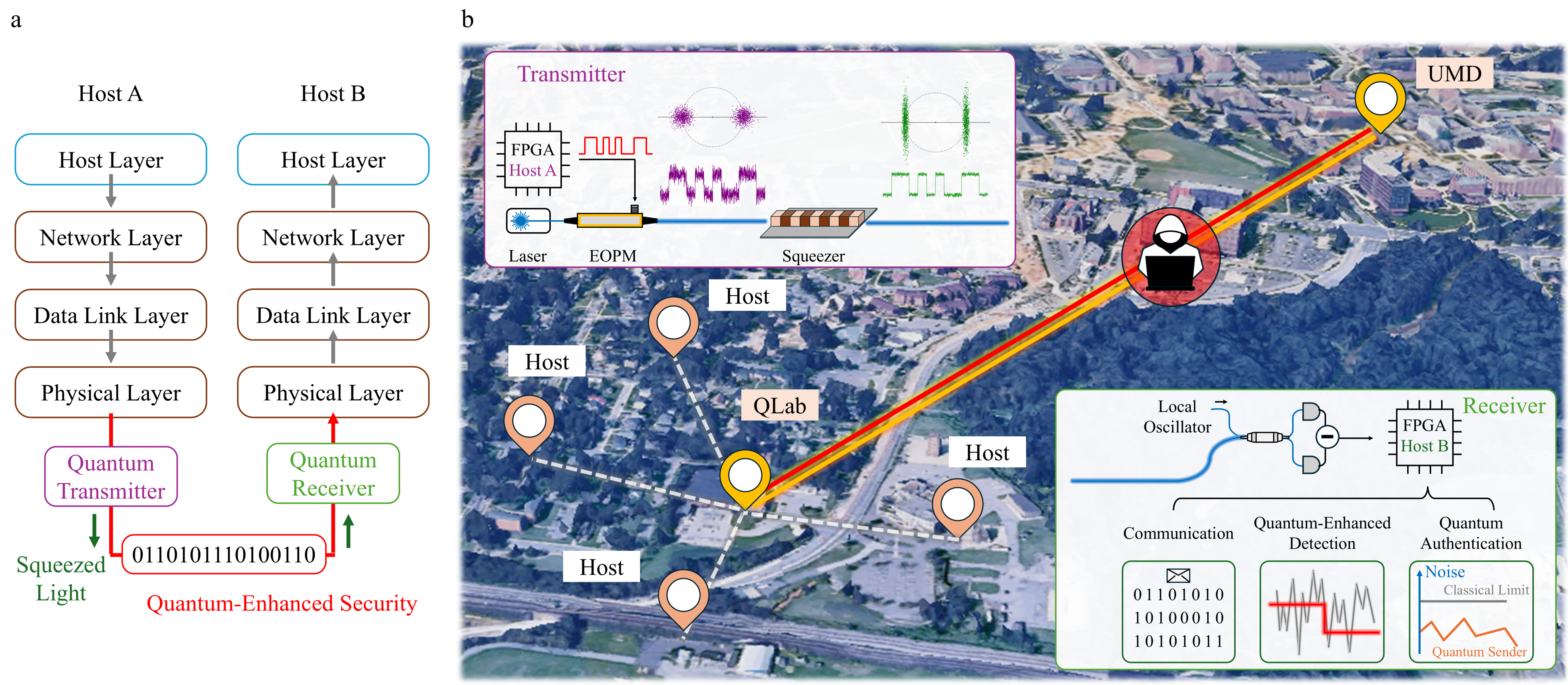}
\caption{
\textbf{Quantum-enhanced optical network}
\textbf{a}, Open system interconnection (OSI) model. A quantum transmitter and receiver are integrated at the physical layer with standard network interfaces.
\textbf{b}, Field test of quantum-enhanced security in metropolitan-scale fiber network. 
Hosts located in QLab communicate through a deployed fiber network connecting QLab and UMD, with an optical switch in QLab enabling path reconfiguration.
Upper left inset, schematic of the quantum transmitter. The signal light is generated by a single-frequency laser operating at around 1550~nm. Bits 0 and 1 are encoded using binary phase modulation with an electro-optic phase modulator (EOPM) driven by an FPGA on host server. The signal then undergoes parametric squeezing in an optical parametric amplifier, where alignment of the squeezing quadrature with the phase-encoding basis enables noise reduction below the shot-noise limit. The noise of the initial coherent state (purple) is suppressed below the shot-noise limit along the encoding quadrature, while preserving phase-encoding information (green). 
Lower right inset, schematic of the quantum receiver. Coherent homodyne detection is employed to measure the incoming optical field. The resulting outputs are then processed by a FPGA on host server to jointly implement communication and quantum sensing tasks. 
%The quantum-enhanced network can improve detection sensitivity and facilitates verification of quantum-capable nodes. 
}
\label{fig1}
\end{figure*}

Optical fiber networks form the backbone of modern communication systems, enabling high-capacity information transfer between distributed nodes~\cite{Agrawal2010,Temprana2015,Agrell2016,Olsson2018,Rademacher2021,Joergensen2022}.Security remains a fundamental and enduring challenge in communication systems~\cite{Gisin2002,Fok2011}.
%In particular, communication systems based on classical laser light are vulnerable to malicious manipulation, as adversaries can exploit comparable optical resources to extract or modify signals without being easily detected at higher network layers.
%To address this limitation, quantum states of light provide a fundamentally different approach to optical communication security. 
By harnessing quantum features such as entanglement and the no‑cloning theorem, we can achieve a security regime fundamentally inaccessible to classical communication~\cite{Gisin2002,Scarani2009,Ekert1991,Bennett1992}.
To date, quantum-enhanced network security has been mostly realized through quantum key distribution (QKD)~\cite{Peev2009,Chen2021,Pittaluga2025,Dynes2019,Sasaki2011,Joshi2020,Wang2014,Gruenenfelder2023,Zahidy2024,Wu2025,Zheng2026}, which fundamentally depends on both quantum randomness at the physical layer and classical post-processing at the information-theoretic layer. This intrinsic dual reliance enforces strong cross-layer interdependency, in sharp contrast to the modular abstractions that underpin conventional communication networks~\cite{Sasaki2018,Mehic2020}. This distinction limits scalability and poses challenges for seamless integration of quantum technology into global communication infrastructure. Here, we introduce and experimentally demonstrate a fundamentally different paradigm for quantum-enhanced communication that overcomes this constraint by preserving modular layer separation with minimal modification to conventional network architectures. By directly synthesizing quantum features and classical data into the same optical signals, our approach localizes the modifications from quantum operations within the physical layer, eliminating the need for classical post-processing at the information-theoretic layer and thereby fully preserving the structure of upper layers in network stacks. Unlike previous quantum–classical co-existing networks that depend on multiplexing independent optical modes—whether temporal, polarization, or frequency~\cite{Wang2024,Berrevoets2022,Thomas2023,Thomas2024}—our quantum‑enhanced communication system eliminates the need for such multiplexing. By embedding quantum features and classical information within the same optical field, the architecture achieves seamless coexistence without introducing additional channel resources or complexity requirements. Furthermore, we validate the practicality of our approach through field deployment over existing metropolitan-scale fiber networks, experimentally confirming both the quantum enhancement and its seamless interoperability with conventional communication infrastructure.

%In particular, squeezed states enable noise reduction below the classical shot-noise limit, offering enhanced sensitivity to optical perturbations. This quantum advantage introduces an additional degree of freedom for simultaneously transmitting information and sensing physical-layer disturbances, thereby enabling integrated communication and security functionalities.
%In this work, we demonstrate a quantum-enhanced optical network based on single-mode squeezed light that exploits these properties for physical-layer threat detection. The system integrates a quantum transmitter and receiver with standard network interfaces, enabling compatibility with conventional communication protocols and extending threat detection capabilities beyond the physical layer.
%We experimentally investigate tapping attacks, where a fraction of the optical power is extracted from the communication channel by an adversary. Such attacks induce measurable changes in both the quadrature amplitudes and noise characteristics of the optical field. These effects are exploited for real-time detection of channel disturbances.
%Finally, we validate the system in a 5 km deployed metropolitan fiber link, demonstrating stable squeezing, reliable communication, and effective attack detection in a realistic environment. These results highlight the potential of quantum-enabled optical networks for secure, resilient, and multi-layer communication infrastructures.

%\LF{We should start with the network layer introduction. This paragraph is only related to the physical layer.}

\begin{figure*}[ht]
\centering
\includegraphics[width=0.98\textwidth]{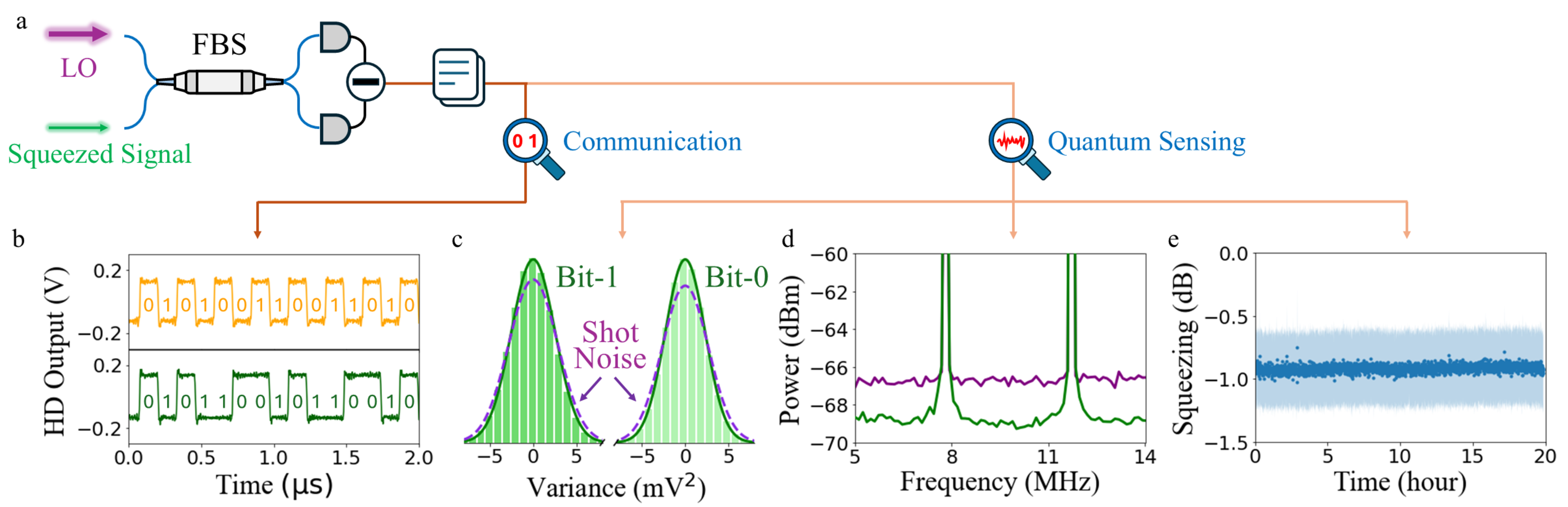}
\caption{
\textbf{Performance of quantum transmitters and receivers.}
\textbf{a}, Schematic of detection and data flow at the quantum receiver.
\textbf{b}, Time-domain traces of decoded signals captured by the analog-to-digital converters (ADCs) with a sampling rate of $\mathrm{1GS/s}$. Idle sequence (orange) and pseudo random binary sequence (green) are shown in the top and bottom plots, respectively. 
%The labeled decoded bits are determined by the two distinct output levels.
\textbf{c}, Statistical distributions of homodyne output for bit-0 and bit-1 in the quantum (green) and classical (purple) cases, based on 80 million samples. The distributions are fitted with Gaussian functions to extract the variances and quantify the squeezing levels.
\textbf{d}, Noise spectra of the homodyne signals for the quantum (green) and classical (purple) cases. The spectra are obtained from the time-domain data via fast Fourier transform (FFT).
\textbf{e}, Squeezing level measurement over 20 hours, demonstrating the stability of the quantum transmitter and receiver.
}
\label{fig2}
\end{figure*}

In modern networks, layered architectures are commonly used to organize data transmission between hosts. Our quantum-enhanced communication systems follow the standard Open Systems Interconnection (OSI) model (Fig.~\ref{fig1}a). The only modification is at the physical layer, where squeezed vacuum is coherently combined with raw classical optical bits. By leveraging the sub-shot-noise variance in squeezed vacuum, we can detect the change in communication channels with enhanced sensitivity beyond the classical limit. 
For data‑link and network‑layer integration, we utilize the Corundum FPGA‑based network interface card (NIC)~\cite{forencich2020fccm}, which provides a high‑performance and programmable interface compatible with conventional networking stacks.
The Corundum NIC further interfaces with the host layer via PCIe, exposing a standard Linux driver and networking API so that host applications can operate without modification.
At the physical layer of the transmitter, the optical field is generated by a single-frequency laser at 1549.6~nm (ITU DWDM D35 channel). After 16-bit digital-to-analog conversion (DAC), raw data bits from the Corundum NIC are encoded in optical fields using a fiber-coupled electro-optic phase modulator (EOPM) with binary phase-shift keying (BPSK) at a clock rate of 7.8 MHz~(Supplementary Section 1).
Modulated optical signals are then injected into an optical parametric amplifier (OPA) for the generation of squeezed light. This enables quantum properties to be directly embedded in optical fields. OPA is realized using a single-pass periodically poled lithium niobate (PPLN) waveguide pumped by the second-harmonic generation (SGH) field (774.8~nm) of the signal optical field (1549.6~nm).
By aligning the BPSK encoding basis with the squeezing quadrature~(See Methods and Supplementary Section 2), the variance of optical signals is reduced below the shot-noise limit (Fig. ~\ref{fig1}b).
At the receiver, coherent homodyne detection is used to extract the encoded classical information and assess the quantum features of optical fields at the physical layer (Fig.~\ref{fig1}c). Analog output from the homodyne detector is digitized by a 1~GHz  16-bit analog-to-digital converter (ADC). 
%Classical data bits and quantum variance are then passed to higher layers separately.
% and quantum variance
Traffic of classical data bits are intercepted by an eXpress Data Path (XDP) program and processed at the driver level, enabling low-latency packet filtering and admission control based on transmitter verification before packets enter the kernel network stack.

%In this work, the quantum transmitter and receiver are directly integrated into the physical layer through standard interfaces with higher network layers to maintain compatibility with existing network protocols.
%Furthermore, the reduced-noise characteristic enhances sensitivity to power variations, while concurrent changes in noise provide a robust signature of channel attenuation.

%At the network level, if a communication path is compromised, the system can dynamically identify and switch to an alternative secure route to maintain continuous communication (Fig.~\ref{fig1}c). 
%Each node comprises a transmitter and a receiver, with communication implemented using the binary phase-shift keying (BPSK) protocol.

We first characterize the physical layer performance of the quantum-enhanced communication system. 
%A phase-modulated signal at 1549.6 nm and a frequency-doubled pump are coupled into a single-pass periodically poled lithium niobate (PPLN) waveguide housed in a temperature-stabilized oven, where bright squeezed light is generated via a $\chi^{(2)}$ optical parametric amplification (OPA) process. 
%The relative phases among the pump, signal, and local oscillator are actively stabilized using two phase-locking loops, enabling reliable retrieval of the encoded information and producing two well-separated output levels corresponding to bit-0 and bit-1 with a clock speed of 7.8~MHz (Fig. \ref{fig2}a).
%\LF{With a transmitted power of 50 nW, raw classical data bits are decoded with a bit-error-rate (BER) below ??? (Fig.~\ref{fig2}a).}
Quantum features of transmitted optical fields are characterized at the same time. Squeezed light can exhibit variance below the classical shot noise limit. This can be revealed with two different approaches in our system. First, we directly calculate the variance of received bit-0 and bit-1 homodyne signals. 
%In parallel, we verify the nonclassical squeezing properties using two independent analyses: variance evaluation of the measured quadratures and noise spectrum analysis. 
The classical baseline is established by turning off the SHG pump for squeezed light generation at the transmitter. The variances of bit-0 and bit-1 signals are measured as $14.36\pm0.003\;\mathrm{mV}^2$ and $16.08\pm0.005\;\mathrm{mV}^2$ respectively (purple in Fig.~\ref{fig2}b). 
%In the regime where the signal is three orders of magnitude weaker than the local oscillator, the measured quadrature variance is dominated by the local oscillator power.
After turning on the SHG pump for squeezed light generation, the variances of bit-0 and bit-1 signals become $10.56\pm0.016\;\mathrm{mV}^2$ and $12.74\pm0.018\;\mathrm{mV}^2$ respectively. This corresponds to a $24\pm0.08\%$ variance reduction or $1.16\pm0.01$~dB below the classical limit (green in Fig.~\ref{fig2}b). 
We can also verify the variance reduction through the noise spectrum, obtained by the Fourier transform of received homodyne signals. Near 10 MHz frequency offset, the noise floor of squeezed light is reduced by $41\pm1.2\%$ or $2.3\pm0.21$~dB below the classical limit (Fig.~\ref{fig2}c).
The higher squeezing level in the noise spectrum is attributed to the narrow-band measurement, in contrast to the effective broadband measurement in the direct calculation of signal variance~(Supplementary Section 3).
%The slightly lower squeezing observed in the variance analysis is attributed to signal power fluctuations.
Long-term stability is critical for the integration of quantum-enhanced communication systems in deployed fiber networks. We perform a continuous squeezing evaluation of our system over 20 hours without human intervention. The squeezing level is maintained over $0.92\pm0.3$~dB below the classical limit throughout the whole test (Fig.~\ref{fig2}d).
%Squeezing below the shot-noise limit is maintained for over 20 hours, demonstrating the suitability of the system for sustained operation in practical environments (Fig.\;\ref{fig2}d).
%These results confirm that quantum-capable nodes can be reliably identified through nonclassical noise characteristics, which cannot be reproduced by classical systems, thereby providing an additional layer of physical-layer security for optical networks.

\begin{figure*}[ht]
\centering
\includegraphics[width=0.98\textwidth]{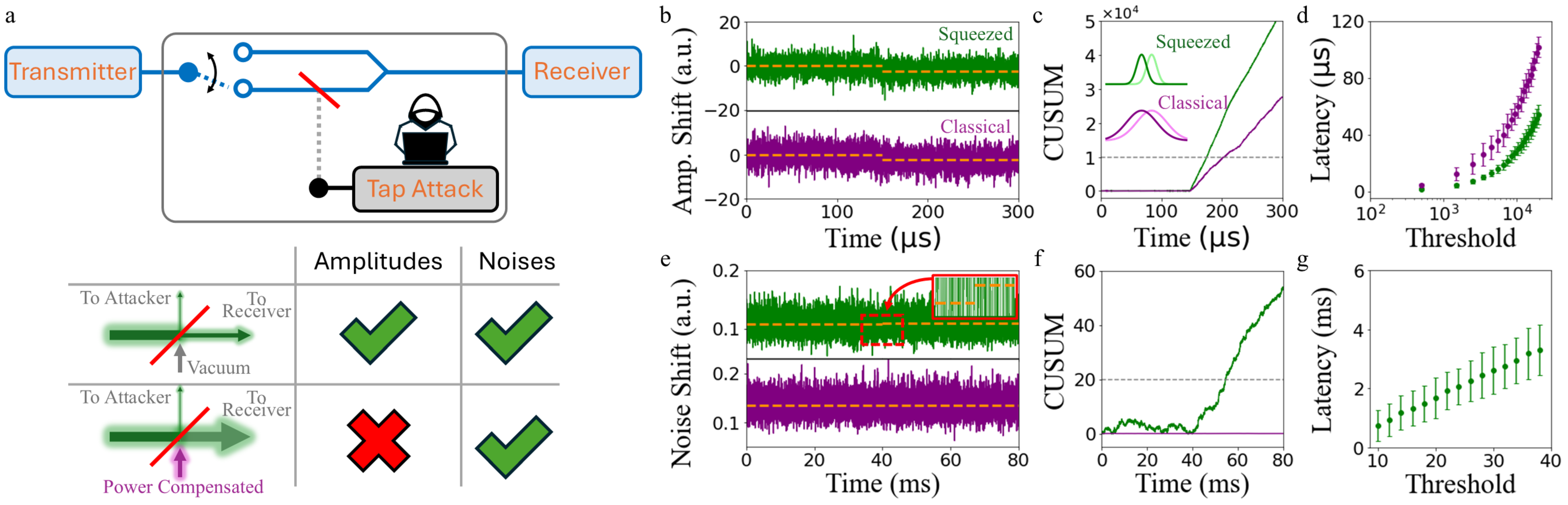}
\caption{
\textbf{Performance of detecting tapping attacks}
\textbf{a}, Schematic of the tapping attack, where an eavesdropper extracts a fraction of the optical power between network nodes. Two representative attack strategies are illustrated. Without power compensation, tapping attacks can be detected by monitoring quadrature amplitudes or noise. 
With power compensation, tapping attacks can only be detected through analyzing the noise property of non-classical signals.
\textbf{b}, Time-domain traces of quadrature amplitudes during the tapping attack test. Mean values are indicated by orange dashed lines.
\textbf{c}, Amplitude-based CUSUM outputs. An alarm is triggered when the output exceeds a predefined threshold (gray line). The quantum case exhibits faster detection due to enhanced sensitivity enabled by squeezed light. The inset illustrates the principle: for the same loss change, reduced variance in the quantum case improves the distinguishability between distributions before and after tapping. 
\textbf{d}, Detection latency as a function of alarm threshold for amplitude-based CUSUM.
\textbf{e}, Time-domain traces of quadrature noise during the tapping attack test. The inset shows a zoomed-in view of the tapping window.
\textbf{f}, Noise-based CUSUM outputs for attack detection. The output for the classical case is always zero.
\textbf{g}, Detection latency as a function of alarm threshold for noise-based CUSUM. The classical noise-based CUSUM does not trigger alarms. 
In \textbf{b}-\textbf{g}, quantum and classical cases are all plotted in green and purple respectively.
}
\label{fig3}
\end{figure*}

Next, we demonstrate the use of squeezed light for the quantum enhancement of communication security.
Our system is evaluated against optical tapping attacks that extract a fraction of the transmitted optical signals at the physical layer (Fig. \ref{fig3}a).
With classical light fields, tapping attacks only change the optical signal amplitude, and can be hidden by coherent power compensation.
With squeezed light, tapping attacks modify both the amplitude and quantum variance of optical signals, and cannot be recovered by classical approaches.
Experimentally, tapping attacks are implemented by an electrically-controlled optical attenuator that can dynamically change the transmission of optical links.
We use cumulative sum (CUSUM) method to detect the small changes in received signals~\cite{Page1954}. The CUSUM output is defined as~(See Methods)
\begin{equation}
\begin{aligned}
\mathrm{S[k]=max(0,log[P_2(x_k)/P_1(x_k)]+S[k-1])},
\label{Eq:CUSUM}
\end{aligned}
\end{equation}
%\SP{[SP: S[k] in CUSUM algorithm is not the running sum since there is a max(0, ) operation.]}
where $S[0] = 0$, $\mathrm{P_{1(2)}}$ denotes the measurement distribution in the absence (presence) of tapping, and $\mathrm{x_k}$ denotes the k-th measurement. An alarm is triggered once the CUSUM output exceeds a predefined threshold.
We first directly use the amplitude of homodyne signals as the CUSUM input ($x_n$ in Eq.~\ref{Eq:CUSUM}). 
%A 5\% tapping is used. 
Under a 5\% tap applied at the middle of the capture,
the reduction in the homodyne signal amplitude is observed in both the classical and quantum cases after tapping (Fig.~\ref{fig3}b). This change is captured by the CUSUM output in both cases (Fig.~\ref{fig3}c).
However, squeezing reduces the measurement noise, thereby increasing the distinguishability between the distributions before and after tapping~\cite{Guha2025,John2025}. This enhanced separation enables the CUSUM algorithm to produce sharper transitions, resulting in a twofold reduction in detection latency compared with the classical case (Fig. \ref{fig3}d). The detection latency is defined as the elapsed time between the onset of the controlled tapping event and the threshold-crossing point of the CUSUM statistic.
% Due to its intrinsic loss dependence, the measured squeezing under tapping can be described as
% \begin{equation}
% \begin{aligned}
% \mathrm{S_{-}=(1-T_0T_{tap})+T_0T_{tap}S_{0}}
% \label{Eq:SQZ}
% \end{aligned}
% \end{equation}
% where $\mathrm{T_0}$ is the system detection efficiency, $\mathrm{(1-T_{tap})}$ is the loss induced by tapping attacks, and $\mathrm{S_0}$ is the squeezing level in the ideal detection system.
%Importantly, if an attacker applies a sophisticated classical power-compensation strategy to maintain the quadrature amplitudes, the measured noise still inevitably increases due to the loss-dependent degradation of squeezing. As a result, the quantum approach outperforms classical methods in detecting tapping attacks, particularly in more general attack scenarios.

\begin{figure*}[ht]
\centering
\includegraphics[width=0.8\textwidth]{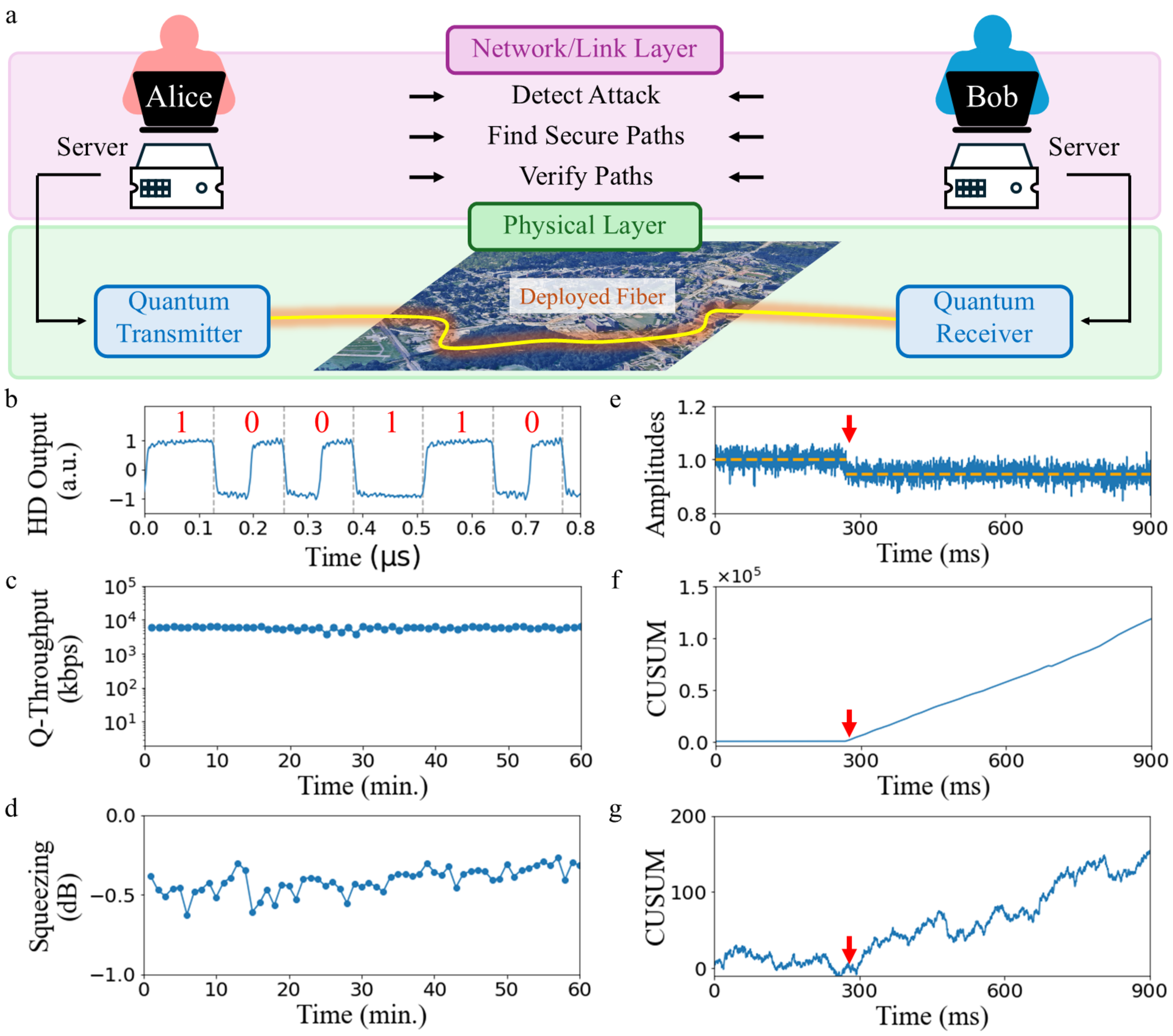}
\caption{
\textbf{Field test of the quantum network}
\textbf{a}, Schematic of the quantum-network field trial. High-layer functionalities are managed by the host server, wheres quantum operation are implemented at the physical layer through a deployed fiber. 
\textbf{b}, Time-domain trace of the received homodyne signal encoded using differential Manchester (DME) after transmission through a 5-km deployed fiber loopback.
\textbf{c}, Measured throughput of the quantum-enhanced channel, characterized using TCP iPerf between Alice and Bob.
\textbf{d}, Measured squeezing level during the field trial.
\textbf{e}, Measured amplitudes trace with a 10~$\%$ tapping event introduced at approximately 300 ms, as indicated by the arrow.
\textbf{f}, Amplitude-based CUSUM output calculated from the amplitude trace shown in \textbf{e}.
\textbf{g}, Noise-based CUSUM output calculated from the same measured record.
}
\label{fig4}
\end{figure*}

We further apply the CUSUM method to analyze the noise spectrum. Specifically, the noise spectral density at 6 MHz is used as the input to the CUSUM algorithm. In the classical case, tapping does not alter the variance of coherent states, and therefore no observable change appears in the noise spectrum, preventing the CUSUM method from detecting the attack. In contrast, for the quantum case, tapping induces an increase in the noise spectrum due to degradation of the squeezing. As a result, the CUSUM analysis reliably identifies the presence of tapping. Notably, this noise-based CUSUM approach with squeezed light remains effective even under ideal classical power compensation, since the observed noise increase is fundamentally linked to the loss-induced degradation of the squeezing level.
%Furthermore, the noise-based CUSUM algorithm also successfully detects the attacks, consistent with amplitude-based approach. Notably, this method has no classical counterpart and remains effective even under ideal power compensation, as the noise increase is intrinsically linked to loss-induced degradation of squeezing (Fig. 3f and 3g). \YC{The detection results can then be forwarded to higher network layers to trigger adaptive operations such as secure-path rerouting.}

After validating quantum-enhanced tapping detection, we integrate the physical layer with higher-layer communication functions and perform field tests over deployed fiber networks (Fig.~\ref{fig4}a). A pair of fibers connecting the National Quantum Laboratory (QLab) and the Atlantic Building at the University of Maryland (UMD) forms a loop exceeding 5 km in length, with a total link attenuation of approximately 9 dB, primarily due to fiber-to-fiber connectors. To mitigate environmental fluctuations in the deployed fibers, differential Manchester encoding (DME) is employed~\cite{Horowitz2015}, in which information is encoded in signal transitions (Fig.~\ref{fig4}b).
%To further demonstrate compatibility with existing optical networks, a high-speed classical channel at 1310 nm is introduced at the physical layer using a pair of quad small form-factor pluggable (QSFP) modules. Using wavelength-division multiplexing (WDM), the quantum signal at 1550 nm and the classical channel at 1310 nm are co-propagated through the same fiber (Fig.~\ref{fig4}a). Then, both channels are processed on servers located at the Alice and Bob sites, respectively, and integrated into higher-layer network functions.
Two hosts, Alice and Bob, are configured as an iPerf client and server, respectively. Alice runs the TCP iPerf client continuously in consecutive 1-minute intervals. Under these conditions, a stable throughput exceeding 6 Mbps is achieved while maintaining 0.5 dB of squeezing (Fig.~\ref{fig4}c and d).
We further evaluate system performance under a tapping attack in the deployed link. A 10\% tapping event is introduced during transmission, and both amplitude- and noise-based CUSUM methods are used for detection. Given that this is a field test, we did not assume that the strength of the tap was known apriori at the receiver. For this purpose, a CUSUM method that does not require knowledge of the post-change distribution was applied (see Eq.~\ref{eq:cusum-mean-based}). For both the amplitude- and noise-based detection approaches, the CUSUM outputs exhibit clear rising trends immediately following the onset of the attack (Fig.~\ref{fig4}e–f), demonstrating the robustness of quantum-enhanced tapping detection under realistic field conditions.
%\SP{[SP: Should we mention that the CUSUM variant being used here doesnt require the knowledge of post-change distribution?]}
%Meanwhile, a coexisting classical communication link simultaneously supports a data rate of 37~Gbps within the same network infrastructure. The preservation of nonclassical noise reduction during data transmission indicates that quantum resources can coexist with practical communication protocols without being completely degraded by environmental perturbations.
%These results demonstrate that quantum-enhanced noise signatures can provide sensitive physical-layer threat detection while maintaining stable network communication.
%More importantly, the proposed scheme combines communication, quantum sensing, and network monitoring within the same optical infrastructure, highlighting the potential of quantum-enhanced approaches for future secure and resilient optical networks.

\begin{figure}[ht]
\centering
\includegraphics[width=0.49\textwidth]{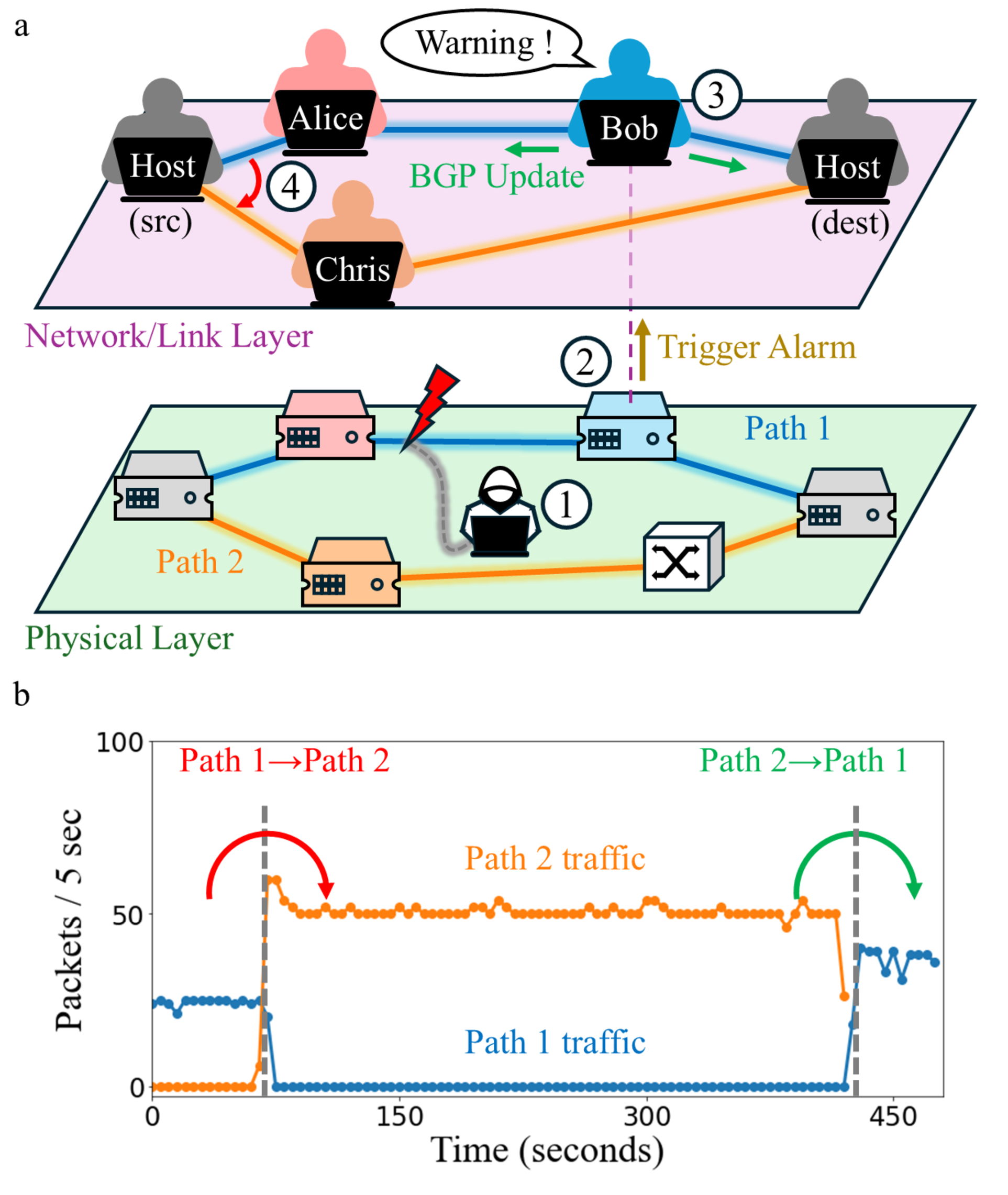}
\caption{
\textbf{Network rerouting in a quantum-enhanced network}
\textbf{a}, Schematic of the rerouting demonstration. Upon threat detection at the physical layer, an alarm is triggered and initiates a BGP update to redirect traffic from the primary path to an alternative secure path in upper layers.
\textbf{b}, Monitored data rate during the rerouting demonstration. Following threat detection, traffic is switched from the primary path (blue) to an alternative secure route (orange), resulting in a corresponding change in the measured data rate. Once the threat is removed, traffic is restored to the original route.
}
\label{fig5}
\end{figure}

Finally, we demonstrate a network-level application enabled by quantum-enhanced tapping detection. Multiple communication paths exist between the source and destination hosts, with the shortest path (Source $\rightarrow$ Alice $\rightarrow$ Bob $\rightarrow$ Destination) initially selected for data transmission (Fig.~\ref{fig5}a). When a tapping attack occurs on the link between Alice and Bob, it is promptly detected by Bob, and an alarm is propagated from the physical layer to higher network layers.
Following detection, Bob announces a Border Gateway Protocol (BGP)~\cite{rfc4271} update to neighboring peers, triggering dynamic rerouting of traffic to an alternate secure path (Fig.~\ref{fig5}b). This adaptive response enables uninterrupted communication without manual intervention. Once the attack is removed, the system automatically restores the original shortest route, demonstrating both the reliability and reconfigurability of the proposed approach for secure optical networking based on quantum-enhanced physical-layer sensing.

In summary, we experimentally demonstrate a quantum-enhanced optical network that integrates communication and physical-layer threat detection using squeezed light. The required modifications to harness quantum resources are fully confined to the physical layer, thereby preserving the modular architecture of conventional network stacks.
By leveraging nonclassical noise signatures, the system enables reliable detection of tapping attacks without disrupting normal data transmission. Beyond threat identification, the network supports adaptive functionalities such as dynamic rerouting, allowing traffic to be automatically redirected to secure paths in response to detected attacks.
These results confirm that quantum resources can provide practical advantages not only for secure information transfer, but also for intelligent monitoring and protection at the physical layer. Our work represents a significant step toward scalable, autonomous quantum-secure optical networks with built-in threat awareness and adaptive protection, offering a promising pathway toward resilient and reconfigurable communication infrastructures enhanced by quantum technologies.

\textbf{Acknowledgments}

This research was supported by the DARPA Quantum Augmented Networking (QuANET) Program under Contract No. HR001124C0405 and partially by the Office of Naval Research (ONR) under Grant No. N00014-25-1-2130, U.S. Department of Energy (Field Work Proposal ERKJ432). L.F. acknowledge the support from Coherent/II-VI Foundation and Sloan Fellowship. This document does not contain technology or Technical Data controlled under either the U.S. International Traffic in Arms Regulations or the U.S. Export Administration Regulations. The authors thanks Michael Grace for discussion.

\vspace{4pt}
\textbf{Author contributions}

The experiments were conceived by Y.C.K, P.B and
L.F. CUSUM algorithm was developed by S.P and P.B. FPGA platform was developed by A.F. Measurements were conducted by Y.C.K, S.P and D.C with supports by C.C, J.P and P.K.C. Data analysis was conducted by Y.C.K., S.P. and D.C.. All authors participated in the manuscript
preparation. N.A, S.G, P.B and L.F supervised the work 

\vspace{4pt}
\textbf{Competing interests}

Authors claim no competing interests.

\textbf{Methods}

\textbf{Phase stabilization}

The relative phases among the signal, pump, and local oscillator must be actively stabilized to ensure stable communication and consistent quadrature measurement. To achieve this, two independent active phase-locking loops are implemented.
The first loop aligns the encoding basis with the squeezing quadrature, while the second loop stabilizes the homodyne measurement basis with respect to the encoded signal. To generate the corresponding error signals, a 62.5 MHz phase dither is applied to the input optical fields.
The error signal for the first loop is extracted by monitoring the tapped output power of the squeezed field, while the second loop error signal is derived from the homodyne detection output. These signals are processed in real time and fed into two independent PID controllers.
Phase stabilization is achieved using fiber stretchers, which compensate for phase drifts in the signal and local oscillator paths, respectively.

\textbf{CUSUM algorithm}

To determine whether a tapping event has occurred from a sequence of measurement results, the cumulative sum (CUSUM) algorithm is employed. The measurement distributions under the no-attack and attack conditions are denoted as $P_1(x)$ and $P_2(x)$ respectively. Based on the Gaussian characterization of the measurement outcomes, these distributions are modeled as,
\begin{equation}
    P_i(x)=\frac{1}{\sqrt{2\pi\sigma_i}}{\exp \left[-\frac{(x-\mu_i)^2}{2\sigma_i^2} \right]}
\end{equation}
where $i\in \{ 1,2\}$, $\mu_i$ and $\sigma_i^2$ represent the mean and variance under each condition.

The log-likelihood ratio for each sample is then given by
\begin{equation}
l[n] = \log \frac{P_2(x_n)}{P_1(x_n)},
\end{equation}
and the CUSUM statistic is recursively updated as
\begin{equation}
S[k] = \max\left(0, S[k-1] + l[k]\right),
\end{equation}
with $S[0]=0$. A detection event is declared when the CUSUM statistic $S[k] > h$, where $h$ is a predefined threshold. 

When the post-change distribution is not known, the log-likelihood based CUSUM cannot be directly applied. For experiments conducted on deployed fibers (Fig.~\ref{fig4}), we used a variant of the CUSUM algorithm that aggregates change in value with respect to the pre-change mean $\mu_1$. Two CUSUM statistics $S^+[k]$ and $S^-[k]$ are maintained that capture positive and negative deviation from the mean respectively. These statistics are recursively updated as 
\begin{align}
S^+[k] &= \max \left( 0, S^+[k] + (x_n - \mu_1) - k  \right), \nonumber \\
S^-[k] &= \max \left( 0, S^-[k] - (x_n - \mu_1) - k  \right),
\label{eq:cusum-mean-based}
\end{align}
with $S^+[0]=0$, $S^-[0]=0$, and $k>0$ representing tolerated deviation in either direction that does not lead to increase in the CUSUM statistics. A positive (negative) change detection event is raised when $S^+[k]>h_1$ ($S^-[k]>h_2$), where $h_1$ and $h_2$ are predefined thresholds. A tap event is detected by the negative CUSUM statistic applied on the amplitude signal and the positive CUSUM statistic applied on the noise spectrum signal.

\bibliography{Ref}

@inproceedings{forencich2020fccm,
    author = {Alex Forencich and Alex C. Snoeren and George Porter and George Papen},
    title = {Corundum: An Open-Source {100-Gbps} {NIC}},
    booktitle = {28th IEEE International Symposium on Field-Programmable Custom Computing Machines},
    year = {2020},
}

@Article{Sasaki2011,
  author    = {Sasaki, M. and Fujiwara, M. and Ishizuka, H. and Klaus, W. and Wakui, K. and Takeoka, M. and Miki, S. and Yamashita, T. and Wang, Z. and Tanaka, A. and Yoshino, K. and Nambu, Y. and Takahashi, S. and Tajima, A. and Tomita, A. and Domeki, T. and Hasegawa, T. and Sakai, Y. and Kobayashi, H. and Asai, T. and Shimizu, K. and Tokura, T. and Tsurumaru, T. and Matsui, M. and Honjo, T. and Tamaki, K. and Takesue, H. and Tokura, Y. and Dynes, J. F. and Dixon, A. R. and Sharpe, A. W. and Yuan, Z. L. and Shields, A. J. and Uchikoga, S. and Legré, M. and Robyr, S. and Trinkler, P. and Monat, L. and Page, J.-B. and Ribordy, G. and Poppe, A. and Allacher, A. and Maurhart, O. and Länger, T. and Peev, M. and Zeilinger, A.},
  journal   = {Optics Express},
  title     = {Field test of quantum key distribution in the Tokyo QKD Network},
  year      = {2011},
  issn      = {1094-4087},
  month     = May,
  number    = {11},
  pages     = {10387},
  volume    = {19},
  doi       = {10.1364/oe.19.010387},
  publisher = {Optica Publishing Group},
}

@Article{Pittaluga2025,
  author    = {Pittaluga, Mirko and Lo, Yuen San and Brzosko, Adam and Woodward, Robert I. and Scalcon, Davide and Winnel, Matthew S. and Roger, Thomas and Dynes, James F. and Owen, Kim A. and Juárez, Sergio and Rydlichowski, Piotr and Vicinanza, Domenico and Roberts, Guy and Shields, Andrew J.},
  journal   = {Nature},
  title     = {Long-distance coherent quantum communications in deployed telecom networks},
  year      = {2025},
  issn      = {1476-4687},
  month     = Apr,
  number    = {8060},
  pages     = {911--917},
  volume    = {640},
  doi       = {10.1038/s41586-025-08801-w},
  publisher = {Springer Science and Business Media LLC},
}

@Article{Chen2021,
  author    = {Chen, Yu-Ao and Zhang, Qiang and Chen, Teng-Yun and Cai, Wen-Qi and Liao, Sheng-Kai and Zhang, Jun and Chen, Kai and Yin, Juan and Ren, Ji-Gang and Chen, Zhu and Han, Sheng-Long and Yu, Qing and Liang, Ken and Zhou, Fei and Yuan, Xiao and Zhao, Mei-Sheng and Wang, Tian-Yin and Jiang, Xiao and Zhang, Liang and Liu, Wei-Yue and Li, Yang and Shen, Qi and Cao, Yuan and Lu, Chao-Yang and Shu, Rong and Wang, Jian-Yu and Li, Li and Liu, Nai-Le and Xu, Feihu and Wang, Xiang-Bin and Peng, Cheng-Zhi and Pan, Jian-Wei},
  journal   = {Nature},
  title     = {An integrated space-to-ground quantum communication network over 4,600 kilometres},
  year      = {2021},
  issn      = {1476-4687},
  month     = Jan,
  number    = {7841},
  pages     = {214--219},
  volume    = {589},
  doi       = {10.1038/s41586-020-03093-8},
  publisher = {Springer Science and Business Media LLC},
}

@Article{Dynes2019,
  author    = {Dynes, J. F. and Wonfor, A. and Tam, W. W. -S. and Sharpe, A. W. and Takahashi, R. and Lucamarini, M. and Plews, A. and Yuan, Z. L. and Dixon, A. R. and Cho, J. and Tanizawa, Y. and Elbers, J. -P. and Greißer, H. and White, I. H. and Penty, R. V. and Shields, A. J.},
  journal   = {npj Quantum Information},
  title     = {Cambridge quantum network},
  year      = {2019},
  issn      = {2056-6387},
  month     = Nov,
  number    = {1},
  volume    = {5},
  doi       = {10.1038/s41534-019-0221-4},
  publisher = {Springer Science and Business Media LLC},
}

@Article{Peev2009,
  author    = {Peev, M and Pacher, C and Alléaume, R and Barreiro, C and Bouda, J and Boxleitner, W and Debuisschert, T and Diamanti, E and Dianati, M and Dynes, J F and Fasel, S and Fossier, S and Fürst, M and Gautier, J-D and Gay, O and Gisin, N and Grangier, P and Happe, A and Hasani, Y and Hentschel, M and Hübel, H and Humer, G and Länger, T and Legré, M and Lieger, R and Lodewyck, J and Lorünser, T and Lütkenhaus, N and Marhold, A and Matyus, T and Maurhart, O and Monat, L and Nauerth, S and Page, J-B and Poppe, A and Querasser, E and Ribordy, G and Robyr, S and Salvail, L and Sharpe, A W and Shields, A J and Stucki, D and Suda, M and Tamas, C and Themel, T and Thew, R T and Thoma, Y and Treiber, A and Trinkler, P and Tualle-Brouri, R and Vannel, F and Walenta, N and Weier, H and Weinfurter, H and Wimberger, I and Yuan, Z L and Zbinden, H and Zeilinger, A},
  journal   = {New Journal of Physics},
  title     = {The SECOQC quantum key distribution network in Vienna},
  year      = {2009},
  issn      = {1367-2630},
  month     = July,
  number    = {7},
  pages     = {075001},
  volume    = {11},
  doi       = {10.1088/1367-2630/11/7/075001},
  publisher = {IOP Publishing},
}

@Book{Agrawal2010,
  author    = {Agrawal, Govind P.},
  publisher = {Wiley},
  title     = {Fiber‐Optic Communication Systems},
  year      = {2010},
  isbn      = {9780470918524},
  month     = Oct,
  doi       = {10.1002/9780470918524},
}

@Article{Gisin2002,
  author    = {Gisin, Nicolas and Ribordy, Grégoire and Tittel, Wolfgang and Zbinden, Hugo},
  journal   = {Reviews of Modern Physics},
  title     = {Quantum cryptography},
  year      = {2002},
  issn      = {1539-0756},
  month     = Mar,
  number    = {1},
  pages     = {145--195},
  volume    = {74},
  doi       = {10.1103/revmodphys.74.145},
  publisher = {American Physical Society (APS)},
}

@Article{Ekert1991,
  author    = {Ekert, Artur K.},
  journal   = {Physical Review Letters},
  title     = {Quantum cryptography based on Bell’s theorem},
  year      = {1991},
  issn      = {0031-9007},
  month     = Aug,
  number    = {6},
  pages     = {661--663},
  volume    = {67},
  doi       = {10.1103/physrevlett.67.661},
  publisher = {American Physical Society (APS)},
}

@Article{Bennett1992,
  author    = {Bennett, Charles H. and Brassard, Gilles and Mermin, N. David},
  journal   = {Physical Review Letters},
  title     = {Quantum cryptography without Bell’s theorem},
  year      = {1992},
  issn      = {0031-9007},
  month     = Feb,
  number    = {5},
  pages     = {557--559},
  volume    = {68},
  doi       = {10.1103/physrevlett.68.557},
  publisher = {American Physical Society (APS)},
}

@Article{Guha2025,
  author    = {Guha, Saikat and John, Tiju Cherian and Gong, Zihao and Basu, Prithwish},
  journal   = {Physical Review Letters},
  title     = {Quantum-Enhanced Quickest Change Detection of Transmission Loss},
  year      = {2025},
  issn      = {1079-7114},
  month     = Nov,
  number    = {21},
  volume    = {135},
  doi       = {10.1103/czg5-3y3d},
  publisher = {American Physical Society (APS)},
}

@Article{Scarani2009,
  author    = {Scarani, Valerio and Bechmann-Pasquinucci, Helle and Cerf, Nicolas J. and Dušek, Miloslav and Lütkenhaus, Norbert and Peev, Momtchil},
  journal   = {Reviews of Modern Physics},
  title     = {The security of practical quantum key distribution},
  year      = {2009},
  issn      = {1539-0756},
  month     = Sept,
  number    = {3},
  pages     = {1301--1350},
  volume    = {81},
  doi       = {10.1103/revmodphys.81.1301},
  publisher = {American Physical Society (APS)},
}

@Article{Joshi2020,
  author    = {Joshi, Siddarth Koduru and Aktas, Djeylan and Wengerowsky, Sören and Lončarić, Martin and Neumann, Sebastian Philipp and Liu, Bo and Scheidl, Thomas and Lorenzo, Guillermo Currás and Samec, Zeljko and Kling, Laurent and Qiu, Alex and Razavi, Mohsen and Stipčević, Mario and Rarity, John G. and Ursin, Rupert},
  journal   = {Science Advances},
  title     = {A trusted node–free eight-user metropolitan quantum communication network},
  year      = {2020},
  issn      = {2375-2548},
  month     = Sept,
  number    = {36},
  volume    = {6},
  doi       = {10.1126/sciadv.aba0959},
  publisher = {American Association for the Advancement of Science (AAAS)},
}

@Article{Wang2014,
  author    = {Wang, Shuang and Chen, Wei and Yin, Zhen-Qiang and Li, Hong-Wei and He, De-Yong and Li, Yu-Hu and Zhou, Zheng and Song, Xiao-Tian and Li, Fang-Yi and Wang, Dong and Chen, Hua and Han, Yun-Guang and Huang, Jing-Zheng and Guo, Jun-Fu and Hao, Peng-Lei and Li, Mo and Zhang, Chun-Mei and Liu, Dong and Liang, Wen-Ye and Miao, Chun-Hua and Wu, Ping and Guo, Guang-Can and Han, Zheng-Fu},
  journal   = {Optics Express},
  title     = {Field and long-term demonstration of a wide area quantum key distribution network},
  year      = {2014},
  issn      = {1094-4087},
  month     = Sept,
  number    = {18},
  pages     = {21739},
  volume    = {22},
  doi       = {10.1364/oe.22.021739},
  publisher = {Optica Publishing Group},
}

@Book{Horowitz2015,
  author    = {Horowitz, Paul and Hill, Winfield},
  publisher = {Cambridge University Press},
  title     = {The art of electronics},
  year      = {2015},
  isbn      = {9780521809269},
}

@Article{Fok2011,
  author    = {Fok, Mable P. and Wang, Zhexing and Deng, Yanhua and Prucnal, Paul R.},
  journal   = {IEEE Transactions on Information Forensics and Security},
  title     = {Optical Layer Security in Fiber-Optic Networks},
  year      = {2011},
  issn      = {1556-6021},
  month     = Sept,
  number    = {3},
  pages     = {725--736},
  volume    = {6},
  doi       = {10.1109/tifs.2011.2141990},
  publisher = {Institute of Electrical and Electronics Engineers (IEEE)},
}

@Article{Gruenenfelder2023,
  author    = {Grünenfelder, Fadri and Boaron, Alberto and Resta, Giovanni V. and Perrenoud, Matthieu and Rusca, Davide and Barreiro, Claudio and Houlmann, Raphaël and Sax, Rebecka and Stasi, Lorenzo and El-Khoury, Sylvain and Hänggi, Esther and Bosshard, Nico and Bussières, Félix and Zbinden, Hugo},
  journal   = {Nature Photonics},
  title     = {Fast single-photon detectors and real-time key distillation enable high secret-key-rate quantum key distribution systems},
  year      = {2023},
  issn      = {1749-4893},
  month     = Mar,
  number    = {5},
  pages     = {422--426},
  volume    = {17},
  doi       = {10.1038/s41566-023-01168-2},
  publisher = {Springer Science and Business Media LLC},
}

@Article{Zahidy2024,
  author    = {Zahidy, Mujtaba and Ribezzo, Domenico and De Lazzari, Claudia and Vagniluca, Ilaria and Biagi, Nicola and Müller, Ronny and Occhipinti, Tommaso and Oxenløwe, Leif K. and Galili, Michael and Hayashi, Tetsuya and Cassioli, Dajana and Mecozzi, Antonio and Antonelli, Cristian and Zavatta, Alessandro and Bacco, Davide},
  journal   = {Nature Communications},
  title     = {Practical high-dimensional quantum key distribution protocol over deployed multicore fiber},
  year      = {2024},
  issn      = {2041-1723},
  month     = Feb,
  number    = {1},
  volume    = {15},
  doi       = {10.1038/s41467-024-45876-x},
  publisher = {Springer Science and Business Media LLC},
}

@Article{Zheng2026,
  author    = {Zheng, Yun and Wang, Hanyu and Jia, Xinyu and Huang, Jiahui and Yuan, Huihong and Zhai, Chonghao and Dai, Junhao and Shi, Jingbo and Zhang, Lei and Zhang, Xuguang and Zhuang, Minxue and Liu, Jinchang and Mao, Jun and Dai, Tianxiang and Fu, Zhaorong and Jiao, Yuqing and Shi, Yaocheng and Dai, Daoxin and Wang, Xingjun and Li, Yan and Gong, Qihuang and Yuan, Zhiliang and Chang, Lin and Wang, Jianwei},
  journal   = {Nature},
  title     = {Large-scale quantum communication networks with integrated photonics},
  year      = {2026},
  issn      = {1476-4687},
  month     = Feb,
  number    = {8104},
  pages     = {68--75},
  volume    = {651},
  doi       = {10.1038/s41586-026-10152-z},
  publisher = {Springer Science and Business Media LLC},
}

@Article{Thomas2023,
  author    = {Thomas, Jordan M. and Kanter, Gregory S. and Kumar, Prem},
  journal   = {Optics Express},
  title     = {Designing noise-robust quantum networks coexisting in the classical fiber infrastructure},
  year      = {2023},
  issn      = {1094-4087},
  month     = Dec,
  number    = {26},
  pages     = {43035},
  volume    = {31},
  doi       = {10.1364/oe.504625},
  publisher = {Optica Publishing Group},
}

@Article{Berrevoets2022,
  author    = {Berrevoets, Remon C. and Middelburg, Thomas and Vermeulen, Raymond F. L. and Chiesa, Luca Della and Broggi, Federico and Piciaccia, Stefano and Pluis, Rene and Umesh, Prathwiraj and Marques, Jorge F. and Tittel, Wolfgang and Slater, Joshua A.},
  journal   = {Communications Physics},
  title     = {Deployed measurement-device independent quantum key distribution and Bell-state measurements coexisting with standard internet data and networking equipment},
  year      = {2022},
  issn      = {2399-3650},
  month     = July,
  number    = {1},
  volume    = {5},
  doi       = {10.1038/s42005-022-00964-6},
  publisher = {Springer Science and Business Media LLC},
}

@Article{Thomas2024,
  author    = {Thomas, Jordan M. and Yeh, Fei I. and Chen, Jim Hao and Mambretti, Joe J. and Kohlert, Scott J. and Kanter, Gregory S. and Kumar, Prem},
  journal   = {Optica},
  title     = {Quantum teleportation coexisting with classical communications in optical fiber},
  year      = {2024},
  issn      = {2334-2536},
  month     = Dec,
  number    = {12},
  pages     = {1700},
  volume    = {11},
  doi       = {10.1364/optica.540362},
  publisher = {Optica Publishing Group},
}

@Article{Wang2024,
  author    = {Wang, Jing and Rollick, Brian J. and Jia, Zhensheng and Huberman, Bernardo A.},
  journal   = {Journal of Lightwave Technology},
  title     = {Time-Interleaved C-Band Co-Propagation of Quantum and Classical Channels},
  year      = {2024},
  issn      = {1558-2213},
  month     = June,
  number    = {11},
  pages     = {4086--4095},
  volume    = {42},
  doi       = {10.1109/jlt.2024.3381105},
  publisher = {Institute of Electrical and Electronics Engineers (IEEE)},
}

@Article{Page1954,
  author    = {Page, E. S.},
  journal   = {Biometrika},
  title     = {Continuous Inspection Schemes},
  year      = {1954},
  issn      = {0006-3444},
  month     = June,
  number    = {1/2},
  pages     = {100},
  volume    = {41},
  doi       = {10.2307/2333009},
  publisher = {JSTOR},
}

@Misc{rfc4271,
  author       = {Yakov Rekhter and Tony Li and Susan Hares},
  howpublished = {RFC 4271},
  title        = {{A Border Gateway Protocol 4 (BGP-4)}},
  year         = {2006},
  publisher    = {Internet Engineering Task Force},
  url          = {https://datatracker.ietf.org/doc/rfc4271/},
}

@InProceedings{John2025,
  author    = {John, Tiju Cherian and Gagatsos, Christos N. and Bash, Boulat A.},
  booktitle = {2025 IEEE Information Theory Workshop (ITW)},
  title     = {Quickest Change-point Detection With Continuous-Variable Quantum States},
  year      = {2025},
  month     = Sept,
  pages     = {1--6},
  publisher = {IEEE},
  doi       = {10.1109/itw62417.2025.11240316},
}

@Article{Mehic2020,
  author    = {Mehic, Miralem and Niemiec, Marcin and Rass, Stefan and Ma, Jiajun and Peev, Momtchil and Aguado, Alejandro and Martin, Vicente and Schauer, Stefan and Poppe, Andreas and Pacher, Christoph and Voznak, Miroslav},
  journal   = {ACM Computing Surveys},
  title     = {Quantum Key Distribution: A Networking Perspective},
  year      = {2020},
  issn      = {1557-7341},
  month     = Sept,
  number    = {5},
  pages     = {1--41},
  volume    = {53},
  doi       = {10.1145/3402192},
  publisher = {Association for Computing Machinery (ACM)},
}

@Article{Wu2025,
  author    = {Wu, Qi and Ribezzo, Domenico and Di Sciullo, Giammarco and Cocchi, Sebastiano and Ann Shaji, Divya and Alves Zischler, Lucas and Luis, Ruben and Serena, Paolo and Lasagni, Chiara and Bononi, Alberto and Hayashi, Tetsuya and Gagliano, Alessandro and Martelli, Paolo and Gatto, Alberto and Parolari, Paola and Boffi, Pierpaolo and Bacco, Davide and Zavatta, Alessandro and Zhu, Yixiao and Hu, Weisheng and Xu, Zhaopeng and Shtaif, Mark and Marotta, Andrea and Graziosi, Fabio and Mecozzi, Antonio and Antonelli, Cristian},
  journal   = {Light: Science \& amp; Applications},
  title     = {Integration of quantum key distribution and high-throughput classical communications in field-deployed multi-core fibers},
  year      = {2025},
  issn      = {2047-7538},
  month     = Aug,
  number    = {1},
  volume    = {14},
  doi       = {10.1038/s41377-025-01982-z},
  publisher = {Springer Science and Business Media LLC},
}

@Article{Sasaki2018,
  author    = {Sasaki, Masahide},
  journal   = {IEEE Security \& amp; Privacy},
  title     = {Quantum Key Distribution and Its Applications},
  year      = {2018},
  issn      = {1558-4046},
  month     = Sept,
  number    = {5},
  pages     = {42--48},
  volume    = {16},
  doi       = {10.1109/msp.2018.3761713},
  publisher = {Institute of Electrical and Electronics Engineers (IEEE)},
}

@Article{Olsson2018,
  author    = {Olsson, Samuel L.I. and Eliasson, Henrik and Astra, Egon and Karlsson, Magnus and Andrekson, Peter A.},
  journal   = {Nature Communications},
  title     = {Long-haul optical transmission link using low-noise phase-sensitive amplifiers},
  year      = {2018},
  issn      = {2041-1723},
  month     = June,
  number    = {1},
  volume    = {9},
  doi       = {10.1038/s41467-018-04956-5},
  publisher = {Springer Science and Business Media LLC},
}

@Article{Rademacher2021,
  author    = {Rademacher, Georg and Puttnam, Benjamin J. and Luís, Ruben S. and Eriksson, Tobias A. and Fontaine, Nicolas K. and Mazur, Mikael and Chen, Haoshuo and Ryf, Roland and Neilson, David T. and Sillard, Pierre and Achten, Frank and Awaji, Yoshinari and Furukawa, Hideaki},
  journal   = {Nature Communications},
  title     = {Peta-bit-per-second optical communications system using a standard cladding diameter 15-mode fiber},
  year      = {2021},
  issn      = {2041-1723},
  month     = July,
  number    = {1},
  volume    = {12},
  doi       = {10.1038/s41467-021-24409-w},
  publisher = {Springer Science and Business Media LLC},
}

@Article{Agrell2016,
  author    = {Agrell, Erik and Karlsson, Magnus and Chraplyvy, A R and Richardson, David J and Krummrich, Peter M and Winzer, Peter and Roberts, Kim and Fischer, Johannes Karl and Savory, Seb J and Eggleton, Benjamin J and Secondini, Marco and Kschischang, Frank R and Lord, Andrew and Prat, Josep and Tomkos, Ioannis and Bowers, John E and Srinivasan, Sudha and Brandt-Pearce, Maïté and Gisin, Nicolas},
  journal   = {Journal of Optics},
  title     = {Roadmap of optical communications},
  year      = {2016},
  issn      = {2040-8986},
  month     = May,
  number    = {6},
  pages     = {063002},
  volume    = {18},
  doi       = {10.1088/2040-8978/18/6/063002},
  publisher = {IOP Publishing},
}

@Article{Temprana2015,
  author    = {Temprana, E. and Myslivets, E. and Kuo, B.P.-P. and Liu, L. and Ataie, V. and Alic, N. and Radic, S.},
  journal   = {Science},
  title     = {Overcoming Kerr-induced capacity limit in optical fiber transmission},
  year      = {2015},
  issn      = {1095-9203},
  month     = June,
  number    = {6242},
  pages     = {1445--1448},
  volume    = {348},
  doi       = {10.1126/science.aab1781},
  publisher = {American Association for the Advancement of Science (AAAS)},
}

@Article{Joergensen2022,
  author    = {Jørgensen, A. A. and Kong, D. and Henriksen, M. R. and Klejs, F. and Ye, Z. and Helgason, O. B. and Hansen, H. E. and Hu, H. and Yankov, M. and Forchhammer, S. and Andrekson, P. and Larsson, A. and Karlsson, M. and Schröder, J. and Sasaki, Y. and Aikawa, K. and Thomsen, J. W. and Morioka, T. and Galili, M. and Torres-Company, V. and Oxenløwe, L. K.},
  journal   = {Nature Photonics},
  title     = {Petabit-per-second data transmission using a chip-scale microcomb ring resonator source},
  year      = {2022},
  issn      = {1749-4893},
  month     = Oct,
  number    = {11},
  pages     = {798--802},
  volume    = {16},
  doi       = {10.1038/s41566-022-01082-z},
  publisher = {Springer Science and Business Media LLC},
}

\end{document}